\documentclass[12pt]{iopart}
\usepackage{iopams}
\usepackage{amssymb}
\usepackage{amsthm}
\usepackage{graphicx}

\newcommand{\bra}[1]{{\left\langle #1 \right|}}
\newcommand{\ket}[1]{{\left| #1 \right\rangle}}
\newtheorem{Thm}{Theorem}

\theoremstyle{definition}

\begin{document}
\title[]{Quantum messages with signatures forgeable in arbitrated quantum signature schemes}

\author{Taewan Kim$^1$, Jeong Woon Choi$^2$, Nam-Su Jho$^3$ and Soojoon Lee$^4$ }

\address{$^1$
 Institute of Mathematical Sciences,
 Ewha Womans University, Seoul 120-750, Korea
}
\address{$^2$
 Fusion Technology R{\&}D Center,
 SK Telecom, Kyunggi 463-784, Korea
}
\address{$^3$
 Cryptography Research Team,
 Electronics and Telecommunications Research Institute,
 Daejeon 305-700, Korea
}
\address{$^4$
 Department of Mathematics and Research Institute for Basic Sciences,
 Kyung Hee University, Seoul 130-701, Korea
}

\eads
{
\mailto{level@khu.ac.kr}
}

\date{\today}

\begin{abstract}
Even though a method to perfectly sign quantum messages has not been known,
the arbitrated quantum signature scheme has been considered as one of good candidates.
However, its forgery problem has been an obstacle to the scheme being a successful method.
In this paper,
we consider one situation, which is slightly different from the forgery problem,
that we check
whether at least one quantum message with signature
can be forged in a given scheme,
although all the messages cannot be forged.
If there exist only a finite number of forgeable quantum messages in the scheme then
the scheme can be secure against the forgery attack
by not sending the forgeable quantum messages,
and so our situation does not directly imply that we check whether the scheme is secure against the attack.
But, if users run a given scheme without any consideration of forgeable quantum messages
then a sender might transmit such forgeable messages to a receiver,
and an attacker can forge the messages if the attacker knows them in such a case.
Thus it is important and necessary to look into forgeable quantum messages.
We here show that there always exists such a forgeable quantum message-signature pair
for every known scheme with quantum encryption and rotation,
and numerically show that
any forgeable quantum message-signature pairs do not exist
in an arbitrated quantum signature scheme.
\end{abstract}

\pacs{
03.67.Dd, 
03.67.Hk 
}
\maketitle

\section{Introduction}
Digital signature has been considered as one of the most important cryptographic tools
for not only authentication of digital messages and data integrity
but also non-repudiation of origin.
Thus, since the advent of quantum cryptography
which provides us with unconditional security in key distribution,
many studies on quantum-mechanics-based signatures have been conducted.

In particular, it was pointed out that
digitally signing quantum messages is not possible~\cite{BCGST}
although quantum mechanics can be helpful in digital signature~\cite{GC}.
Hence, quantumly signing quantum messages with the help of an arbitrator
has been suggested~\cite{ZK,LCL,CM,ZQ,GQGW,CCH,LLZC,ZZL,SL,ZQSSS,ZLS},
and the signature schemes are called
the {\em arbitrated quantum signature} (AQS) schemes.

In most AQS schemes on the qubit system, their quantum signature operators consist of two parts.
One is called the random rotation $\{R_j\}_{j\in\mathbb{Z}_2}$ defined by
two Pauli operators $\sigma_x$ and $\sigma_z$, that is,
$R_{0}=\sigma_x$ and $R_{1}=\sigma_z$,
and the other is called the quantum encryption $\{E_k\}_{k\in\mathbb{Z}_4}$~\cite{BR}
such that for all qubit states $\rho$
\begin{equation}
\frac{1}{4}\sum_{k\in\mathbb{Z}_4}E_{k}\rho E_{k}^\dagger = \frac{1}{2}I,
\label{eq:E_k}
\end{equation}
where $E_k$ are unitary operators.

In the AQS schemes,
by applying these two parts of operators to a given quantum message $\ket{M}$
according to the previously shared key $(j,k)$,
the signature
\begin{equation}
\ket{S}=E_k R_j \ket{M}
\label{eq:Sign}
\end{equation}
is obtained,
and the validity of the signature can basically be determined as follows:
Let $\ket{M'}$ be the transmitted message and $R_j^{\dagger}E_k^{\dagger}\ket{S'}$
be the state obtained
by applying the inverse of quantum signature operators
to the transmitted signature $\ket{S'}$,
then the signature is valid if and only if
\begin{equation}
\ket{M'}\simeq R_j^{\dagger}E_k^{\dagger}\ket{S'},
\label{eq:valid}
\end{equation}
where $A\simeq B$ means that $A$ and $B$ are the same up to global phase.
In other words, for each $j\in\mathbb{Z}_2$ and $k\in\mathbb{Z}_4$,
there exists a real number $\theta_{jk}$ such that
\begin{equation}
\ket{M'}=e^{i\theta_{jk}} R_j^{\dagger}E_k^{\dagger}\ket{S'}.
\label{eq:valid_gp}
\end{equation}
We note that
one can judge with high probability
whether or not the two states $\ket{M'}$ and $R_j^{\dagger}E_k^{\dagger}\ket{S'}$
are equal up to global phase,
by exploiting the swap test~\cite{BCWW} for appropriate number of copies of the states.

We remark that all quantum encryptions are not useful for AQS schemes.
In particular, it has been shown that if quantum encryption
consists of only the Pauli operators $\sigma_x$, $\sigma_y$, $\sigma_z$ and the identity operator $I$,
then the AQS schemes with the quantum encryption are not secure
against a receiver's forgery attack~\cite{GQGW,CCH,ZZL,ZQSSS,ZLS}.
In order to recover the security of the AQS schemes,
the following form of quantum encryption $E_k$ was proposed~\cite{CCH}:
For $k\in \mathbb{Z}_4$,
$E_k=V\sigma_k W$,
where $V$ and $W$ are proper unitary operators,
$\sigma_{0}=I$, $\sigma_{1}=\sigma_{x}$, $\sigma_{2}=\sigma_{y}$ and $\sigma_{3}=\sigma_{z}$.
However, if the above encryption is employed then, as seen in Eq.~(\ref{eq:Sign}),
the unitary operator $V$ in the signature
$\ket{S}=V\sigma_k W R_j \ket{M}$
can always be eliminated by an attacker's applying the inverse of $V$.
Therefore, the quantum encryption proposed in Ref.~\cite{CCH}
can be reduced to the encryption
\begin{equation}
E_k =\sigma_k W,
\label{eq:q_encryption}
\end{equation}
for $k\in \mathbb{Z}_4$.
This unitary operator $W$ is called
an {\em assistant unitary operator} of the AQS scheme~\cite{ZLS}.

Let us consider a situation that there exists a non-identity unitary operator $Q$
such that all the operators $R_j^{\dagger}W^{\dagger}\sigma_k Q\sigma_k W R_j$
become the identical unitary operator $U$ up to global phase,
regardless of the shared key $(j,k)$,
that is, for all $j\in\mathbb{Z}_2$ and $k\in\mathbb{Z}_4$,
\begin{equation}
R_j^{\dagger}W^{\dagger}\sigma_k Q\sigma_k W R_j\simeq U.
\label{eq:forgery}
\end{equation}
We remark that if $\ket{S}=\sigma_k W R_j \ket{M}$ and the transmitted message-signature pair
is $(\ket{M},\ket{S})$
then the pair can be forged as $(U\ket{M},Q\ket{S})$
since the forged message $U\ket{M}$ and the forged signature $Q\ket{S}$
satisfy the validity condition~(\ref{eq:valid}), that is,
for all $j\in\mathbb{Z}_2$ and $k\in\mathbb{Z}_4$
\begin{equation}
U\ket{M}\simeq R_j^{\dagger}W^{\dagger}\sigma_k Q\ket{S}.
\label{eq:forgery_message}
\end{equation}
It follows that it is possible
for the receiver to forge all quantum message-signature pairs
in this situation,
and it can hence be shown that
the scheme with the quantum encryption and the rotation satisfying Eq.~(\ref{eq:forgery})
is insecure against a forgery attack.

Recently, Zhang {\em et al.}~\cite{ZLS}
pointed out that
if an unitary operator $Q$ satisfies Eq.~(\ref{eq:forgery}) for some unitary $U$ and $W$
then $Q$ must be one of the Pauli operators.
Furthermore, for each Pauli operator $\sigma_l$,
they characterised the class of the assistant unitary operators $W$
satisfying the following:
There exists an unitary $U$
such that
\begin{equation}
R_j^{\dagger}W^{\dagger}\sigma_k \sigma_l\sigma_k W R_j\simeq U
\label{eq:forgery_sigma_l}
\end{equation}
for all $j\in\mathbb{Z}_2$ and $k\in\mathbb{Z}_4$.
From the characterisation, one can obtain the class of the $W$'s
that provide us with quantum encryptions
in which all quantum message-signature pairs cannot be forged.
As an example of such a secure assistant unitary operator,
$W_a$ was introduced in Ref.~\cite{ZLS}, where
\begin{equation}
W_a=\frac{1}{\sqrt{2}}
\left(
\begin{array}{cc}
1 &  e^{i\pi/4} \\
e^{-i\pi/4} & -1 \\
\end{array}
\right),
\label{eq:W_a}
\end{equation}
and it was shown that there is no unitary $U$ which satisfies Eq.~(\ref{eq:forgery_sigma_l}).

Now, let us take into account a slightly different situation as the following:
For a given assistant unitary operator $W$,
there exist a quantum message $\ket{M_0}$,
non-identity unitary $Q$ and unitary $U$
such that
\begin{equation}
R_j^{\dagger}W^{\dagger}\sigma_k^{\dagger}Q\sigma_{k}WR_{j}\ket{M_0}\simeq U\ket{M_0}
\label{eq:forgeable}
\end{equation}
for all $j\in\mathbb{Z}_2$ and $k\in\mathbb{Z}_4$.
This implies that
the receiver can forge at least one quantum message and its signature
although all other quantum message-signature pairs cannot be forged.
Here,
a quantum message satisfying Eq.~(\ref{eq:forgeable})
is said to be {\em forgeable} in the AQS scheme with an assistant unitary operator $W$.
For example, if the operator $W_a$ in Eq.~(\ref{eq:W_a}) is given
as an assistant unitary operator in the AQS scheme then
the computational basis states $\ket{c}$ become forgeable quantum messages,
since
\begin{equation}
R_j^{\dagger}W_a^{\dagger}\sigma_k^{\dagger}\sigma_3\sigma_{k}W_a R_{j}\ket{c}
\simeq \sigma_1\ket{c}
\label{eq:forgeable_W_a}
\end{equation}
for all $j\in\mathbb{Z}_2$ and $k\in\mathbb{Z}_4$,
where $c$ is 0 or 1.
Hence, even though there does not exist any unitary $U$ satisfying Eq.~(\ref{eq:forgery_sigma_l})
in the AQS scheme with $W_a$ as an assistant unitary operator,
there can exist a forgeable quantum message in the AQS scheme.

We note that, assuming that
there are only a finite number of forgeable quantum messages in a given scheme,
then all other messages except the forgeable messages have no problem to be transmitted,
and thus it is possible to secure the scheme from the forgery attack
if the forgeable messages are not sent to the receiver.
Therefore, the scheme can be secure against the forgery attack,
although there exist forgeable quantum messages in the scheme.
However, if we do not consider forgeable quantum messages in a given scheme
then users might use such forgeable messages in the scheme,
and the messages can be forged if the attacker knows them in this situation.
Hence forgeable quantum messages should be investigated and analysed in studying AQS schemes.

In this paper, we show that
for every known AQS scheme with the random rotation $\{R_j\}_{j\in\mathbb{Z}_2}$
and the quantum encryption $\{\sigma_k W\}_{k\in\mathbb{Z}_4}$
as in Eq.~(\ref{eq:q_encryption}),
there always exists at least one forgeable quantum message.
In this situation,
one question naturally arises, such as
whether there exists an AQS scheme without any forgeable quantum message.
In this paper, we numerically show that
there exists no forgeable quantum message
in an AQS scheme with proper random rotation and quantum encryption.

\section{Forgeable messages in AQS schemes}\label{subsec:forgery}
For any unitary $W$, we note that
\begin{eqnarray}
\sigma_{1}W^{\dagger}\sigma_{1}W\sigma_{1}
&\simeq&\sigma_{1}W^{\dagger}\sigma_{k}\sigma_{1}\sigma_{k}W\sigma_{1},\nonumber\\
\sigma_{3}W^{\dagger}\sigma_{1}W\sigma_{3}
&\simeq&\sigma_{3}W^{\dagger}\sigma_{k}\sigma_{1}\sigma_{k}W\sigma_{3},
\label{eq:simeq_prop}
\end{eqnarray}
for all $k\in\mathbb{Z}_4$,
since $\sigma_1$ commutes or anti-commutes with all Pauli matrices, that is,
$\sigma_{1}\simeq \sigma_{k}\sigma_{1}\sigma_{k}$ for all $k\in\mathbb{Z}_4$.
Thus it follows that
there exists a forgeable message $\ket{M_0}$
with the forgery unitary operators $Q=\sigma_1$ and
$U\simeq\sigma_{1}W^{\dagger}\sigma_{1}W\sigma_{1}$ or
$\sigma_{3}W^{\dagger}\sigma_{1}W\sigma_{3}$
in an AQS scheme with
the random rotation $\{R_j\}_{j\in\mathbb{Z}_2}$
and a quantum encryption $\{\sigma_k W\}_{k\in\mathbb{Z}_4}$
if there exists a message $\ket{M_0}$ such that
\begin{equation}
\sigma_{1}W^{\dagger}\sigma_{1}W\sigma_{1}\ket{M_0}
\simeq\sigma_{3}W^{\dagger}\sigma_{1}W\sigma_{3}\ket{M_0}.
\label{eq:simeq_prop2}
\end{equation}
In particular, Eq.~(\ref{eq:simeq_prop2})
is essentially equivalent to the statement that
$\ket{M_0}$ is an eigenstate of the unitary operator
\begin{equation}
\sigma_{3}W^{\dagger}\sigma_{1}W\sigma_{3}\sigma_{1}W^{\dagger}\sigma_{1}W\sigma_{1}
\simeq\sigma_{3}W\sigma_{1}W^{\dagger}\sigma_{2}W^{\dagger}\sigma_{1}W\sigma_{1}
\label{eq:sigma_312}
\end{equation}
with eigenvalue $e^{i\theta}$ for some real number $\theta$.

However, since any unitary operator is normal and its eigenvalues have all modulus one,
there always exists such an eigenstate of the unitary operator in Eq.~(\ref{eq:sigma_312})
by the spectral decomposition theorem.
Similarly, it can be also shown that there exists a forgeable quantum message
with respect to the forgery unitary operators $Q=\sigma_l$ and $U\simeq\sigma_{1}W^{\dagger}\sigma_{l}W\sigma_{1}$,
where $l= 2, 3$.
This implies that the following theorem holds.

\begin{Thm}\label{Thm1}
Assume that an AQS scheme consists of
the random rotation $\{R_j\}_{j\in\mathbb{Z}_2}$
and a quantum encryption $\{\sigma_k W\}_{k\in\mathbb{Z}_4}$
with an assistant unitary operator $W$.
Then there exists at least one forgeable qubit message $\ket{M_0}$,
that is,
there exist a qubit message $\ket{M_0}$ and forgery unitary operators $Q$ and $U$
satisfying Eq.~(\ref{eq:forgeable}) for all $j\in\mathbb{Z}_2$ and $k\in\mathbb{Z}_4$.
\end{Thm}

We remark that
Theorem~\ref{Thm1} can be also shown by a constructive proof.
In other words,
in a given AQS scheme with the random rotation $\{R_j\}_{j\in\mathbb{Z}_2}$
and a quantum encryption $\{\sigma_k W\}_{k\in\mathbb{Z}_4}$,
one can find a forgeable message $\ket{M_0}$ and forgery unitary operators $Q=\sigma_1$ and $U$.
For example, for the AQS scheme with the assistant unitary operator $W_a$ in Eq.~(\ref{eq:W_a}),
one can construct a forgeable quantum message
\begin{equation}
\ket{M_0}=\frac{1}{\sqrt{2}\sqrt{3-\sqrt{3}}}\left(\left(\sqrt{3}-1\right)\ket{0}+\sqrt{2}\ket{1}\right)
\label{eq:W_a_forgeable}
\end{equation}
and forgery unitary operators $Q=\sigma_1$ and
$U\simeq W_a$ or $W^*_a$,
and can also show that, for all $j\in\mathbb{Z}_2$ and $k\in\mathbb{Z}_4$,
\begin{eqnarray}
R_j^{\dagger}W^{\dagger}&\sigma_k&\sigma_1\sigma_k W R_j\ket{M_0}
\nonumber\\
&&\simeq
\frac{1}{\sqrt{2}\sqrt{3-\sqrt{3}}}\left(\sqrt{2}e^{i\pi/6}\ket{0}+\left(\sqrt{3}-1\right)e^{-5i\pi/6}\ket{1}\right)
\nonumber\\
&&\simeq W_a\ket{M_0}~\mathrm{or}~W_a^*\ket{M_0}.
\label{eq:special_forgery}
\end{eqnarray}
In general, for an AQS scheme with an arbitrary assistant unitary operator $W$,
a forgeable quantum message can be constructed as follows.

Without loss of generality,
we may assume that an assistant unitary operator $W$ has the following representation~\cite{unitary}:
\begin{eqnarray}
W &=& w_0 \sigma_0+iw_1 \sigma_1-iw_2 \sigma_2 +iw_3 \sigma_3,
\label{eq:W}
\end{eqnarray}
where $w_j \in\mathbb{R}$, $w_0\ge 0$
and $\sum_{j\in\mathbb{Z}_4}w_j^2 =1$.
Let
\begin{eqnarray}
\alpha
&=&\frac{1}{2}\left(w_{0}^{2}+w_{1}^{2}-w_{2}^{2}-w_{3}^{2}\right)
=w_{0}^{2}+w_{1}^{2}-\frac{1}{2}
=\frac{1}{2}-w_{2}^{2}-w_{3}^{2},
\nonumber \\
\beta&=&w_{0}w_{2}+w_{1}w_{3},
\nonumber \\
\gamma&=&w_0w_3-w_1w_2.
\label{eq:alphabetagamma}
\end{eqnarray}
If $\beta=0$ then it can readily be obtained that
\begin{eqnarray}
\sigma_{1}W^{\dagger}\sigma_{1}W\sigma_{1}\ket{0}
&=& 2\left(\alpha-i\gamma\right)\ket{1}
\nonumber \\
&\simeq&-2\left(\alpha+i\gamma\right)\ket{1}
\nonumber \\
&=& \sigma_{3}W^{\dagger}\sigma_{1}W\sigma_{3}\ket{0},
\label{eq:beta0}
\end{eqnarray}
which implies
\begin{equation}
R_j^{\dagger}W^{\dagger}\sigma_k \sigma_1\sigma_k W R_j\ket{0}
\simeq \sigma_{1}W^{\dagger}\sigma_{1}W\sigma_{1}\ket{0},
\label{eq:beta0_forgery}
\end{equation}
for all $j\in\mathbb{Z}_2$ and $k\in\mathbb{Z}_4$,
since it is clear that
\begin{eqnarray}
\sigma_{1}W^{\dagger}\sigma_{1}W\sigma_{1}
&\simeq&\sigma_{1}W^{\dagger}\sigma_{k}\sigma_{1}\sigma_{k}W\sigma_{1},\nonumber\\
\sigma_{3}W^{\dagger}\sigma_{1}W\sigma_{3}
&\simeq&\sigma_{3}W^{\dagger}\sigma_{k}\sigma_{1}\sigma_{k}W\sigma_{3},
\label{eq:simeq_prop}
\end{eqnarray}
for all $k\in\mathbb{Z}_4$.
Since if we take $Q=\sigma_1$ and $U=\sigma_{1}W^{\dagger}\sigma_{1}W\sigma_{1}$
then Eq.~(\ref{eq:beta0_forgery}) is equivalent to
the forgeability condition in Eq.~(\ref{eq:forgeable}),
we can say that the qubit message $\ket{0}$ is forgeable
in AQS schemes with the random rotation $\{R_j\}_{j\in\mathbb{Z}_2}$
and a quantum encryption $\{\sigma_k W\}_{k\in\mathbb{Z}_4}$
whose assistant unitary operator $W$ satisfies $\beta=0$.

We now assume that $\beta\neq 0$,
and let $\ket{M_0}$ be a qubit message defined as
\begin{equation}
\ket{M_0}=\frac{1}{\sqrt{\mu^2+1}}\left(\mu\ket{0}+\ket{1}\right),
\label{eq:M0}
\end{equation}
where
\begin{eqnarray}
\mu=\frac{\alpha+\sqrt{\alpha^2+\beta^2}}{\beta},
\label{eq:mu}
\end{eqnarray}
and let $Q$ and $U$ be forgery unitary operators,
which are defined as $Q=\sigma_1$ and
\begin{equation}
U=2\left(
    \begin{array}{cc}
      -\beta & \alpha+i\gamma \\
      \alpha-i\gamma & \beta \\
    \end{array}
  \right)
 ~~\mathrm{or}~~
  2\left(
    \begin{array}{cc}
       \beta & -\alpha+i\gamma \\
      -\alpha-i\gamma & -\beta \\
    \end{array}
  \right).
  \label{eq:general_U}
\end{equation}
Then it follows that, for all $j\in\mathbb{Z}_2$ and $k\in\mathbb{Z}_4$,
\begin{eqnarray}
R_j^{\dagger}W^{\dagger}&\sigma_k&\sigma_1\sigma_k W R_j\ket{M_0}
\nonumber\\
&&\simeq
\frac{\sqrt{2\beta}(\alpha-\beta\mu+\gamma i)}
{\sqrt{\alpha\mu+\beta}}\ket{0}
+\frac{\sqrt{2}((\alpha-\gamma i)\beta\mu+\beta^2)}
{\sqrt{(\alpha\mu+\beta)\beta}}\ket{1}
\nonumber\\
&&\simeq U\ket{M_0}.
\label{eq:general_forgery}
\end{eqnarray}
Hence one can construct a forgeable quantum message
in an arbitrary AQS scheme of the form in Theorem~\ref{Thm1}.

\section{AQS schemes without forgeable messages}\label{sec:unforgeable}
We have shown that, for every assistant unitary operator $W$,
there exists at least one forgeable qubit message in the AQS scheme
with the random rotation $\{R_j\}_{j\in\mathbb{Z}_2}$
and a quantum encryption $\{\sigma_k W\}_{k\in\mathbb{Z}_4}$.
In this section, we numerically show that
any forgeable qubit message does not exist
in an AQS scheme with a slightly modified random rotation
and a suitable assistant unitary operator.

In order to get rid of forgeable quantum messages,
we first point out that the random rotation $\{R_j\}_{j\in\mathbb{Z}_2}$
in the known AQS schemes is biased,
and the biased random rotation may be one of reasons
why there exists a forgeable quantum message.
Thus we here use an AQS scheme
with an unbiased random rotation $\{\tilde{R}_j\}_{j\in\mathbb{Z}_4}$,
where $\tilde{R}_j=\sigma_j$ for each $j\in\mathbb{Z}_4$.
We note that the random rotation $\{\tilde{R}_j\}_{j\in\mathbb{Z}_4}$
is a kind of quantum encryption
since the random rotation satisfies Eq.~(\ref{eq:E_k}),
and so the above scheme can be considered as
one of AQS schemes with sequential quantum encryption,
which were presented in Ref.~\cite{ZQSSS}.

We now find an assistant unitary operator which we can consider as
one of the most suitable ones.
In order to find it,
we begin with observing one simple case,
such as the case that all quantum messages are forgeable
in a given AQS scheme with its random rotation $\{\tilde{R}_j\}_{j\in\mathbb{Z}_4}$
and quantum encryption $\{\sigma_k W\}_{k\in\mathbb{Z}_4}$.

In particular, Table~\ref{table} shows us what assistant unitary operators $W$
can make all qubit messages forgeable
in the AQS scheme,
when a forgery attack operator $Q$,
which is in Eq.~(\ref{eq:forgery}),
is one of the Pauli matrices.
For example,
if $Q=\sigma_1$ then all qubit messages become forgeable
when the operator $W$ satisfies two of the three equations,
$\alpha=0$, $\beta=0$ and $\gamma=0$,
since
\begin{eqnarray}
\alpha &=& w_0^2 + w_1^2 - \frac{1}{2},\nonumber\\
\beta &=& w_0 w_2 + w_1 w_3,\nonumber\\
\gamma &=& w_0 w_3 - w_1 w_2,
\label{eq:general_forgeable_W}
\end{eqnarray}
as seen in Eqs.~(\ref{eq:alphabetagamma}).

\begin{table}
\begin{center}
\begin{tabular}{c|c}
  \hline
  \hline
  $Q$ & $W=w_0\sigma_0+iw_1 \sigma_1-iw_2 \sigma_2 +iw_3 \sigma_3$ \\
  \hline
             & $w_0^2 + w_1^2 - 1/2 = w_0 w_3 - w_1 w_2 = 0$\\
  $\sigma_1$ & $w_0^2 + w_1^2 - 1/2 = w_0 w_2 + w_1 w_3 = 0$\\
             & $w_0 w_3 - w_1 w_2 = w_0 w_2 + w_1 w_3 = 0$\\
  \hline
             & $w_0^2 + w_2^2 - 1/2 = w_0 w_1 -w_2 w_3 = 0$\\
  $\sigma_2$ & $w_0^2 + w_2^2 - 1/2 = w_0 w_3 +w_1 w_2 = 0$\\
             & $w_0 w_1 - w_2 w_3 = w_0 w_3 +w_1 w_2 = 0$\\
  \hline
             & $w_0^2 + w_3^2 - 1/2 = w_0 w_2 - w_1 w_3 = 0$\\
  $\sigma_3$ & $w_0^2 + w_3^2 - 1/2 = w_0 w_1 + w_2 w_3 = 0$\\
             & $w_0 w_2 - w_1 w_3 = w_0 w_1 + w_2 w_3 = 0$\\
  \hline
  \hline
\end{tabular}
\caption{Characaterisation of assistant unitary operators $W$
which make all qubit messages forgeable in AQS schemes with
its random rotation $\{\tilde{R}_j\}_{j\in\mathbb{Z}_4}$ and
quantum encryption $\{\sigma_k W\}_{k\in\mathbb{Z}_4}$
when a given forgery attack $Q$ is one of the Pauli matrices:
For each forgery attack $\sigma_l$,
if an assistant unitary operator $W$ satisfies one of the three pairs of equations in $w_j$'s
then all qubit messages become forgeable.
}
\label{table}
\end{center}
\end{table}
If an assistant unitary operator $W$ has at most two non-zero $w_j$'s, then
such an operator $W$ satisfies at least one of nine pairs of equations in $w_j$'s
which appear in Table~\ref{table},
and thus all qubit messages are forgeable in the AQS scheme.
Hence we note that at least three $w_j$'s should be non-zero,
in order for all qubit messages not to be forgeable.
However, all unitary operator with at least three non-zero $w_j$'s
are not good candidates of assistant operators
for AQS schemes without forgeable quantum messages.
For examples,
since
\begin{equation}
W_a \simeq \frac{i}{2}\left(\sigma_1-\sigma_2 +\sqrt{2}\sigma_3\right),
\label{eq:W_a_our_form}
\end{equation}
$W_a$ has three non-zero $w_j$'s.
Nevertheless, in the AQS scheme with the assistant unitary operator $W_a$ in Eq.~(\ref{eq:W_a_our_form}),
the computational basis states $\ket{c}$ become forgeable quantum messages
since for all $j\in\mathbb{Z}_4$ and $k\in\mathbb{Z}_4$,
\begin{equation}
\tilde{R}_j^{\dagger}W_a^{\dagger}\sigma_k^{\dagger}\sigma_3\sigma_{k}W_a \tilde{R}_{j}\ket{c}
\simeq \ket{c\oplus 1} = \sigma_1\ket{c},
\label{eq:forgeable_W_a}
\end{equation}
where $\oplus$ is the addition modulo 2.

Here we take an operator $T$ defined by
\begin{equation}
T=\frac{i}{\sqrt{3}}\left(\sigma_{1}-\sigma_{2}+\sigma_{3}\right)
\label{eq:T}
\end{equation}
as an assistant unitary operator for an AQS scheme without forgeable messages,
and  let $d(\cdot,\cdot)$ be a distance between two unitary operators
defined as
\begin{equation}
d(\Phi,\Psi)=|\phi_0-\psi_0|+|\phi_1-\psi_1|+|\phi_2-\psi_2|+|\phi_3-\psi_3|,
\label{eq:distance}
\end{equation}
where $\Phi$ and $\Psi$ are unitary operators with form in Eq.~(\ref{eq:W}), that is,
\begin{eqnarray}
\Phi &\simeq& \phi_0 \sigma_0+i\phi_1 \sigma_1-i\phi_2 \sigma_2 +i\phi_3 \sigma_3, \nonumber\\
\Psi &\simeq& \psi_0 \sigma_0+i\psi_1 \sigma_1-i\psi_2 \sigma_2 +i\psi_3 \sigma_3.
\label{eq:Phi_Psi}
\end{eqnarray}
Then the operator $T$ among unitary operators of form in Eq.~(\ref{eq:W})
with at least three non-zero coefficients of $\sigma_j$'s
is one of the farthest ones from the identity operator $\sigma_0$
with respect to the distance defined in Eq.~(\ref{eq:distance}),
that is, for any unitary $W$,
\begin{equation}
d(\sigma_0, W) \le 1+\sqrt{3} = d(\sigma_0,T).
\end{equation}
Therefore, the operator $T$ could be considered as one good candidate
of an assistant unitary operator for an AQS scheme without forgeable messages.

\begin{figure}
\includegraphics[width=\linewidth]{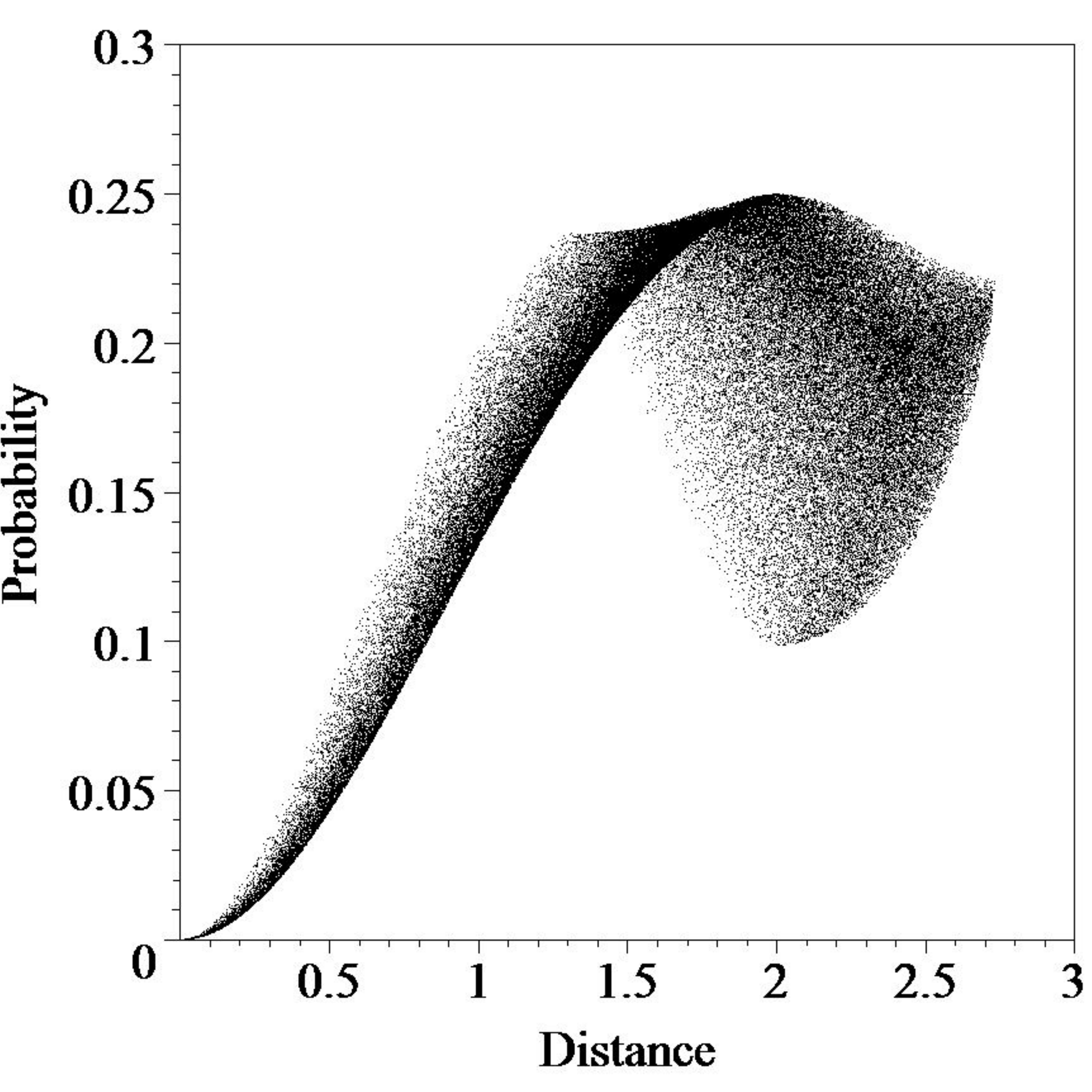}
\caption{\label{Fig:Swaptest}
The minimal probability to detect the forgery attack over all qubit messages
by exploiting the swap test once,
plotted against the distance defined in Eq.~(\ref{eq:distance})
from the identity operator $\sigma_0$
for 100,000 randomly chosen unitary operators $Q$ in Eq.~(\ref{eq:Q}):
When the operator is one of the Pauli matrices (the distance is 2),
the minimal probability to detect a forgery attack, $P_{\min}$,
has a local minimum.
}
\end{figure}

From now on, we numerically investigate the forgeability of the AQS scheme with
the random rotation $\{\tilde{R}_j\}_{j\in\mathbb{Z}_4}$
and the quantum encryption $\{\sigma_k T\}_{k\in\mathbb{Z}_4}$.
Let $Q$ be an arbitrary forgery attack operator defined as
\begin{equation}
Q \simeq q_0\sigma_0 + i q_1\sigma_1 - i q_2\sigma_2 + i q_3\sigma_3,
\label{eq:Q}
\end{equation}
where $q_j$ are real numbers with $q_0\ge 0$ and $\sum_{j\in\mathbb{Z}_4} q_j^2 =1$,
and let $d_Q$ be its distance from the identity operator $\sigma_0$,
that is,
\begin{equation}
d_Q = d(\sigma_0,Q) = |1-q_0| + |q_1| + |q_2| + |q_3|.
\label{eq:dQ}
\end{equation}
For each qubit message $\ket{M}$,
let $P_{Q,\ket{M}}$ be the probability with which
a forgery attack can be detected by using the swap test once,
then it follows from Ref.~\cite{BCWW} that
\begin{eqnarray}
P_{Q,\ket{M}} = 1 - \frac{1}{2^9}\sum_{j,k,j',k'\in\mathbb{Z}_4}
\left(1+|\bra{M}\Delta_{jkj'k'}\ket{M}|^2\right),
\label{eq:PdQ}
\end{eqnarray}
where
\begin{equation}
\Delta_{jkj'k'} = \sigma_{j}T^{\dagger}\sigma_{k}Q^\dagger\sigma_{k}T\sigma_{j}
\sigma_{j'}T^{\dagger}\sigma_{k'}Q\sigma_{k'}T\sigma_{j'}.
\label{eq:Sjk}
\end{equation}
Let $P_Q$ be the minimum of $P_{Q,\ket{M}}$
taken over all qubit messages $\ket{M}$,
then the value of $P_Q$ can efficiently be calculated for a given $Q$.
For 100,000 randomly chosen $Q$,
the points $(d_Q,P_Q)$ are plotted in Figure~\ref{Fig:Swaptest}.

For each $0\le d\le 1+\sqrt{3}$,
let $P_{\min}(d)$ be the minimum of $P_{Q}$'s
taken over all unitary operators $Q$ with $d_Q=d$,
then $P_{\min}(d)$ can be described
as the greatest lower bound of the points $(d_Q,P_Q)$
in Figure~\ref{Fig:Swaptest},
from which we can furthermore see
that $d=0$ if and only if $P_{\min}(d)=0$,
that is, a forgery attack operator is not the identity operator up to global phase
if and only if its detection probability is strictly positive.
This directly implies that
there does not exist any forgeable messages in this AQS scheme.

In addition, we can see from Figure~\ref{Fig:Swaptest}
that, for a forgery attack operator with distance less than $3/2$,
the minimal probability to detect the attack is small
if and only if the operator is close to the identity operator.
Therefore, we can obtain that
the maximal probability not to detect a forgery attack,
$\left(1-P_{\min}(d)\right)^n$, is exponentially close to zero
by performing sufficiently large number $n$ of swap tests
for $n+1$ copies of the message-signature pairs,
and hence one can detect any forgery attack
with arbitrarily small error probability
in the AQS scheme with $\{\tilde{R}_j\}_{j\in\mathbb{Z}_4}$ and $\{\sigma_k T\}_{k\in\mathbb{Z}_4}$
as a random rotation and a quantum encryption, respectively.

\section{Conclusion}\label{sec:Conclusion}
We have considered forgeable quantum messages in AQS schemes,
and have shown that
there exists at least one forgeable quantum message-signature pair
for almost all known AQS schemes.
Finally, we have numerically shown that
there does not exist any forgeable quantum messages
in the AQS scheme with sequential quantum encryptions
$\{\tilde{R}_j\}_{j\in\mathbb{Z}_4}$ and  $\{\sigma_k T\}_{k\in\mathbb{Z}_4}$.
Moreover,
it can be shown that
the arbitrator can confirm the fact that
a sender signed the message
since the information of the sender's secret key
is involved in the signature,
and hence it is impossible
for a sender to disavow the signature~\cite{ZK,LCL,ZQ}.

However, since this scheme uses more random rotation operators than the previous ones,
it needs users' more key strings shared in advance,
and plenty of copies of the message-signature pairs should be required,
in order to detect a forgery attack operator
quite close to the identity operator.
This means that the AQS scheme demands quite a few both classical and quantum resources.
In addition, the AQS scheme may have other security problems
such as the information leakage from many copies of the messages,
which has not been analysed in this paper.
Hence, we cannot say that the AQS scheme is practically useful.

Nevertheless, we can still say that
it is helpful to study AQS schemes without forgeable quantum messages
in improving theoretical works related to AQS,
since the forgeability may invoke another problem which we have not dealt with in AQS.
Therefore, our result could be a basic reference
for both theoretical and practical applications of AQS,
such as finding a practically useful AQS scheme without forgeable messages,
and would also be helpful to strengthen theories in quantum cryptography.

\section*{Acknowledgments}
This work was supported by the Next-Generation Information Computing Development Program 
through the National Research Foundation of Korea (NRF) funded by 
the Ministry of Science, ICT and Future Planning (Grant No.2011-0029925). 
TK was supported by Basic Science Research Program through the National Research Foundation of Korea (NRF) 
funded by the Ministry of Education (2009-0093827), 
and JWC was supported by the ICT R\&D program of MSIP/IITP 
(10044559, Development of key technologies for quantum cryptography network). 
SL acknowledges Gerardo Adesso's hospitality 
during a long-term visit to Quantum Correlations Group 
in the University of Nottingham.



\begin{thebibliography}{00}
\bibitem{unitary}
Any $2\times 2$ unitary operator can be expressed as
the form of Eq.~(\ref{eq:W}) up to global phase.
\end{thebibliography}
\end{document}